\documentclass[]{article}
\usepackage[utf8]{inputenc}
\usepackage{textcomp}
\usepackage[margin=1in]{geometry}
\usepackage{setspace}
\usepackage[parfill]{parskip}
\usepackage{graphicx}
\usepackage[numbers]{natbib}
\usepackage{amsmath}
\usepackage{subcaption}
\usepackage{float}
\usepackage{authblk}
\usepackage[superscript]{cite}
\usepackage{soul}

\title{
Modeling Polaron Excitations and Stabilization Mechanisms \\
in Conjugated Polymers}
\author{Vishal Jindal, Scott T. Milner$^*$}
\affil{Department of Chemical Engineering, The Pennsylvania State University, \\ University Park, PA, 16802, United States}
\affil{stm9@psu.edu}
\date{}

\begin{document}

\maketitle

\begin{abstract}
    Charge carriers in organic semiconductors form polarons,
    which are self-localized states stabilized by interactions with their environment. 
    Using a dielectric-stabilized tight-binding model parameterized 
    from first-principles calculations, 
    we compute ground and excited polaron states in poly(3-hexylthiophene) (P3HT). 
    Our results quantitatively reproduce key mid-infrared absorption features, notably the chain-length-dependent shift of the intrachain polaron excitation peak (peak B) and its variation between regioregular and regiorandom P3HT. 
    Comparison to an alternative stabilization mechanism based on local ring distortions reveals that dielectric polarization dominates polaron formation, as ring distortions yield insufficient binding and excitation energies inconsistent with experiments. 
    These findings clarify the microscopic origin of polarons in conjugated polymers and provide a predictive framework linking polaron energetics to spectroscopic observables, 
    advancing the understanding of charge transport in organic semiconductors.
\end{abstract}

\section{Introduction}


In organic semiconductors, 
a charge carrier (electron or hole) tends to form a polaron. 
A polaron results when a charge carrier interacts with 
its environment to produce a self-bound state in which the carrier 
is stabilized by the deformation of the surrounding medium. \cite{Ghosh2020ExcitonsMaterials}
The two basic mechanisms are moving partial charges and deforming of electron clouds.
The former involves moving the positions of atoms on which partial charges resides,
and the latter involves distorting the electron clouds that form chemical bonds.
Together, these local and electronic distortions create a self-consistent 
potential well that localizes the carrier and defines the polaron state.

Polarons are important in organic semiconductors because hopping is 
only strong in one dimension and the disorder messes up delocalization.
Unlike organic crystals with three dimensional intermolecular coupling,
the conductivity in semiconducting polymers is only one-dimension 
with weak connections across chains which led to charges being 
localized into one-dimensional polarons.
Polaron hopping mechanisms have been used to model charge transport 
in disordered organic polymers such as amorphous poly(3-hexyl)thiophene (P3HT).

While the direct experimental access to polaron size, shape, 
or mobility remain challenging, mid-infrared absorption spectroscopy 
provides a window into the energy landscape of polaronic states.
Specifically, it probes optical transitions between ground and 
excited polaron configurations. \cite{Osterbacka2000a, Ghosh2019a, Kahmann2018}
In poly(3-hexylthiophene) (P3HT), 
the mid-infrared absorption spectrum (see Fig.\ 1) reveal two prominent features: 
a narrow, low-energy peak (labeled as peak A) 
and a broader, higher-energy peak (peak B). 
Peak A appears relatively insensitive to chain morphology, 
while peak B shifts significantly between regioregular (RR) and regiorandom (RRa) P3HT, suggesting a dependence on the underlying polymer structure.

\begin{figure}[h!]
    \centering
    \includegraphics[width=0.75\linewidth]{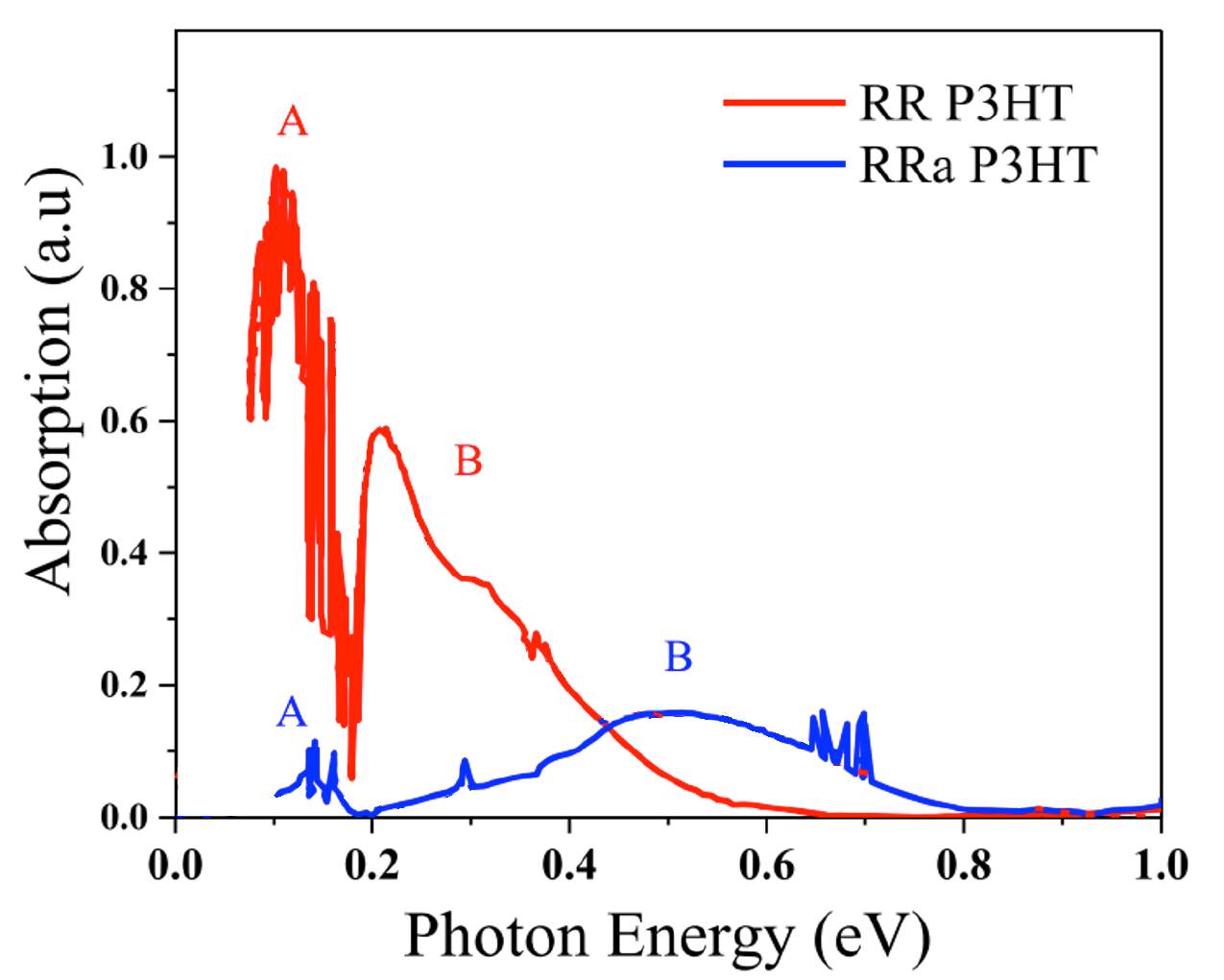}
    \caption{
    Experimental mid-infrared absorption spectra of regioregular (RR) 
    and regiorandom (RRa) poly(3-hexylthiophene) (P3HT). 
    Two distinct absorption features are observed: 
    Peak A, a narrow low-energy band that appears in both RR and RRa P3HT, 
    and Peak B, a broader high-energy band whose position 
    and intensity vary significantly between RR and RRa-P3HT.}
    \label{fig:ExpPolAbsSpec}
\end{figure}

These spectral features reflect optical transitions between polaronic states, 
but they do not directly measure the spatial extent or mobility of the polaron. 
Interpreting these transitions requires a theoretical framework that 
connects the observed energies to specific electronic configurations. 
In this work, we treat the polaron as a bound state 
whose ground and excited wavefunctions can be computed 
using a tight-binding model incorporating dielectric stabilization. 
Within this model, the energy difference between ground 
and excited states corresponds to the observed absorption peaks, 
and the spatial profile of the wavefunction,
which we refer to as delocalization, affects the transition energy.

Previous models of polaron absorption spectra in conjugated polymers 
have often relied on phenomenological assumptions, 
introducing adjustable parameters to fit 
experimental spectra without microscopic justification. 
Many models described polaron absorption by introducing 
localized in-gap states stabilized by lattice distortions, 
with their energies and spatial extent tuned empirically 
to reproduce the observed two-peak mid-infrared structure. 
\cite{Ghosh2016PolaronFilms, Ghosh2018SpectralExperiment}
To account for variations between ordered and disordered films, 
these models invoked empirical distributions of 
conjugation lengths or energetic disorders \cite{Pochas2014}, 
or treated the absorption features as collective plasmonic modes 
with parameters adjusted to match mobility values.
\cite{Rohnacher2021, Spano2010TheJ-aggregates}
While such approaches captured broad spectral features, 
they lacked predictive power and did not clarify the microscopic origin 
of polaron stabilization or excitation energetics. 

Here, we apply a previously established tight-binding model incorporating
dielectric stabilization to investigate polaron excitations in conjugated polymers. 
\cite{Bombile2017} 
The model accounts for both kinetic delocalization along the conjugated backbone 
and electrostatic stabilization from induced polarization 
in the surrounding dielectric medium.
All model parameters including onsite energies, hopping matrix elements, 
and dielectric polarization energies are either extracted or calculated
from first-principles electronic structure calculations 
using density functional theory (DFT). 
 
We use this framework to interpret the shift in peak B 
as a consequence of changes in intrachain polaron structure. 
Regioregular P3HT, with its planar backbone and extended conjugation, 
supports broader polaron wavefunctions and lower excitation energies. 
In contrast, regiorandom P3HT introduces torsional disorder that 
restricts conjugation length, leading to more localized polarons 
and higher excitation energies. 
By comparing model predictions to experimental spectra, 
we aim to evaluate whether dielectric stabilization 
alone can account for these trends, 
or whether local vibrational modes such as 
local ring distortions must be considered.



This paper is structured as follows.
First, we present the dielectric-stabilized tight-binding model and 
its parameterization scheme using DFT. 
Next, we compute polaron ground and excited states across varying chain lengths, 
and compare the predicted excitation energies with polaron absorption spectrum peak B. 
To assess the possible role of local geometric distortions, 
we formulate a model of ring-coupled polarons based on quantum chemical calculations. 
Finally, we analyze and contrast the contributions of both stabilization mechanisms, 
providing a quantitative framework for understanding polaron formation in conjugated polymers.

\section{Tight-binding polaron model}

A charge carrier on a conjugated polymer chain induces 
dipoles in the surrounding polarizable medium. 
The attractive interaction between the charge and 
its induced polarization lowers the system's total energy, 
leading to a stabilized, self-localized polaron state. 
This stabilization arises from electronic polarization 
of nearby molecules and the polymer environment, 
distinct from local geometric distortions of the polymer backbone.

The total polaron energy $E_{\mathrm{Pol}}$ is expressed as the sum of two terms:
\begin{equation}
E_{\mathrm{Pol}} = E_K + E_P,
\end{equation}
where $E_K$ is the electronic kinetic energy of the delocalized carrier along the chain, 
and $E_P$ is the polarization energy arising from dielectric stabilization.

The kinetic energy is described using a one-dimensional tight-binding Hamiltonian:
\begin{equation}
H_{\mathrm{TB}} = - t \sum_{k=1}^{n-1} \left( c_k^\dagger c_{k+1} + c_{k+1}^\dagger c_k \right).
\end{equation}
where $c_k^\dagger$ and $c_k$ are creation and annihilation operators on site $k$,
and $t$ is the nearest-neighbor hopping matrix element, 
extracted from DFT calculations on thiophene monomer and dimer. 
For a planar P3HT chain, we obtain $t = 1.23$ eV for hole polarons. \cite{Jindal2024a}

The tight-binding wavefunction is written as $\psi_a = \sum_i a_i |i\rangle$, 
where $|a_i|^2$ is the probability of finding the charge on site $i$. 
The kinetic energy is then: \begin{equation}
E_K = -t \sum_{i} \left( a_i^* a_{i+1} + a_{i+1}^* a_i \right)
\end{equation}

To compute the polarization energy, 
we treat the surrounding environment as a linear dielectric continuum 
with relative permittivity $\epsilon_r$. 
The charge density on each site is approximated as a three-dimensional 
Gaussian distribution with width $\sigma_i$. 
This smeared-charge representation avoids the divergence 
of the Coulomb self-interaction for point charges. 
The Gaussian width is extracted from DFT calculations of 
the charge distribution on monomer units, 
yielding $\sigma_i = 2 \rm\AA$ for thiophene monomers.

The unscreened Coulomb interaction between two Gaussian charge distributions 
on sites $i$ and $j$ is:
\begin{equation}
E_C(i,j) = \frac{1}{|r_i - r_j|} \, \mathrm{erf}\left( \frac{|r_i - r_j|}{2 \sigma_{\mathrm{avg}}} \right)
\end{equation}
where $\sigma_{\mathrm{avg}} = (\sigma_i + \sigma_j)/2$. The total Coulomb self-energy of the charge distribution is then:
\begin{equation}
E_C = \sum_{i,j} |a_i|^2 |a_j|^2 E_C(i,j)
\end{equation}
The polarization energy is computed from:
\begin{equation}
E_P = -\frac{1}{2} \left( 1 - \frac{1}{\epsilon_r} \right) \sum_{i,j} |a_i|^2 |a_j|^2 E_C(i,j)
\end{equation}

For P3HT, the electronic component of dielectric constant is taken as $\epsilon_r = 2.5$, 
obtained by fitting the optical dispersion model
to the measured complex refractive index. \cite{Hughes2018DeterminingSpectroscopy}
The dielectric-stabilized tight-binding model and parameterization scheme
are described in detail in our previous works. \cite{Bombile2017}.

\begin{figure}[h]
    \centering
    \includegraphics[width=0.75\linewidth]{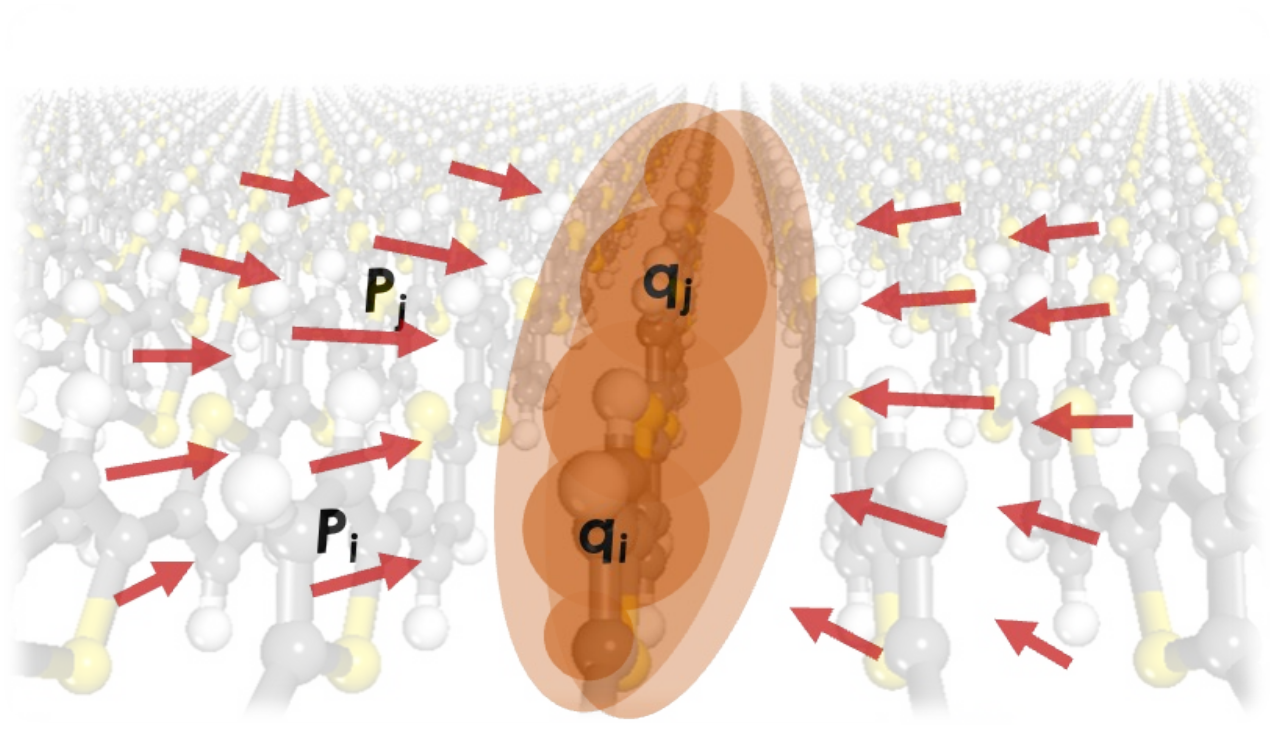}
    \caption{The charge carrier and its induced polarization in the 
            surrounding medium modeled as smeared charge distributions.}
    \label{fig:Polarization}
\end{figure}

Physically, the polarization energy arises from two effects, 
as shown in Fig.\ \ref{fig:Polarization}:
\begin{itemize}
\item[(i)] The charge carrier induces a polarization charge density $\rho_{\mathrm{ind}}(\vec{r}) = \{P_i \}$ in the surrounding dielectric.
\item[(ii)] This induced polarization interacts attractively with the original charge density $\rho(\vec{r}) = \{q_i\}$, lowering the total energy.
\end{itemize}
By modeling both $\rho$ and $\rho_{\mathrm{ind}}$ as distributed Gaussians, 
we capture the long-range, nonlocal stabilization accurately 
without introducing divergences at short range.

The total polaron energy $E_{\mathrm{Pol}}$ is minimized 
with respect to the wavefunction amplitudes $\{a_i\}$, 
under the normalization constraint $\sum_i |a_i|^2 = 1$. 
This self-consistent minimization balances the delocalization 
favored by the kinetic energy $E_K$ with 
the localization driven by the polarization energy $E_P$. 
The resulting ground-state polaron is localized 
over 4–5 monomer units for long P3HT chains.

\section{Polaron excitation}

The excited states of polarons play a key role in their optical properties and charge transport. 
By calculating the first excited state and its energy difference from the ground state, 
we can directly connect theory to experimental absorption spectra 
and gain insights into polaron localization and stabilization.

\begin{figure}[h!]
    \centering
    \includegraphics[width=0.8\linewidth]{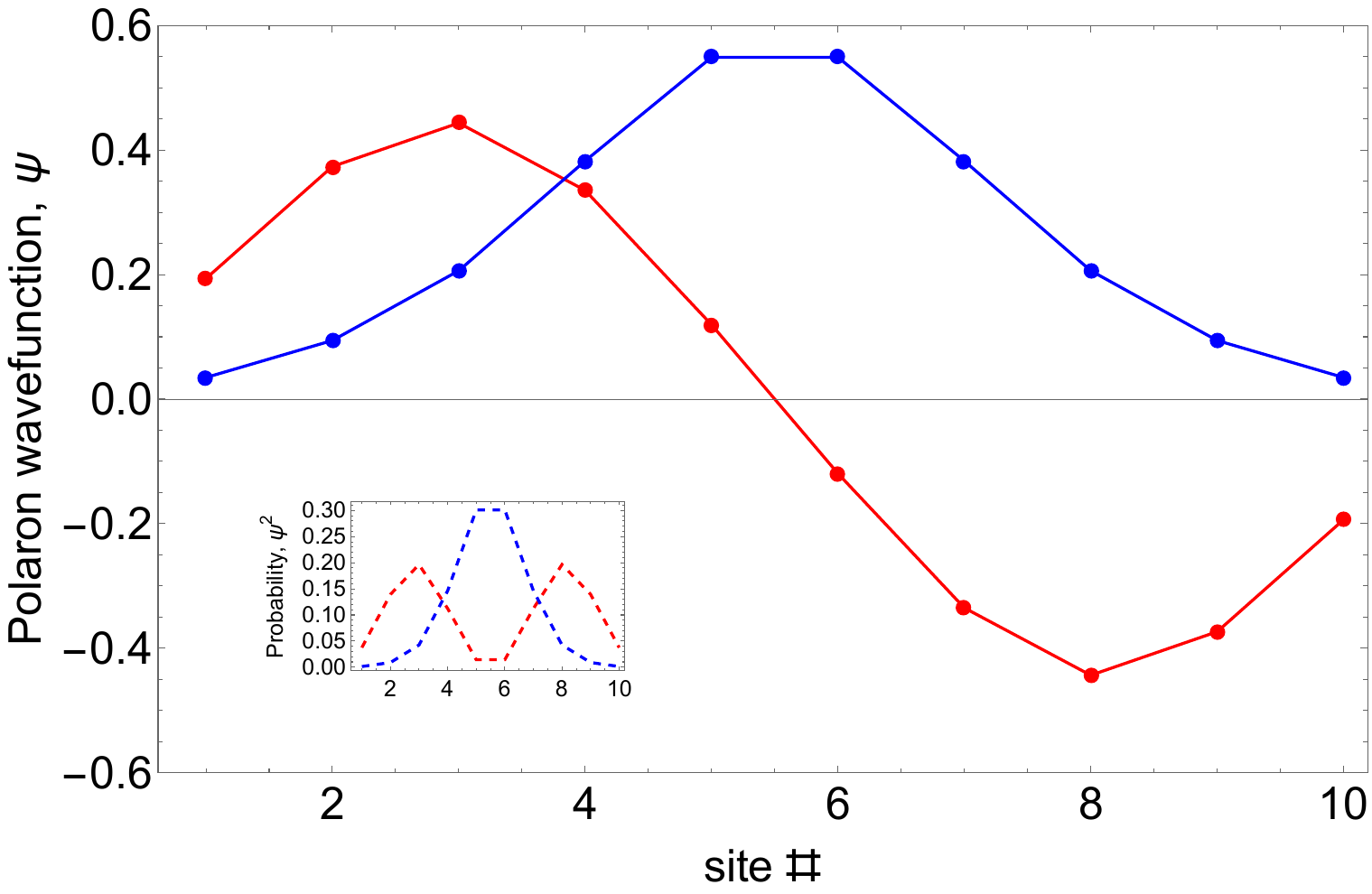}
    \caption{Ground state (blue) and excited state (red) polaron wavefunctions $\Psi$ on straight P3HT chain consisting of 10 monomers. 
    The inset shows the probability distribution $\Psi^2$ for the same polaron wavefunction.}
    \label{fig:ExcitedPolaron_10}
\end{figure}

The first excited state is obtained by imposing 
antisymmetry relative to the chain center. 
The energy difference between the excited and 
ground states defines the polaron excitation energy.

\begin{figure}[h!]
    \centering
    \includegraphics[width=0.8\linewidth]{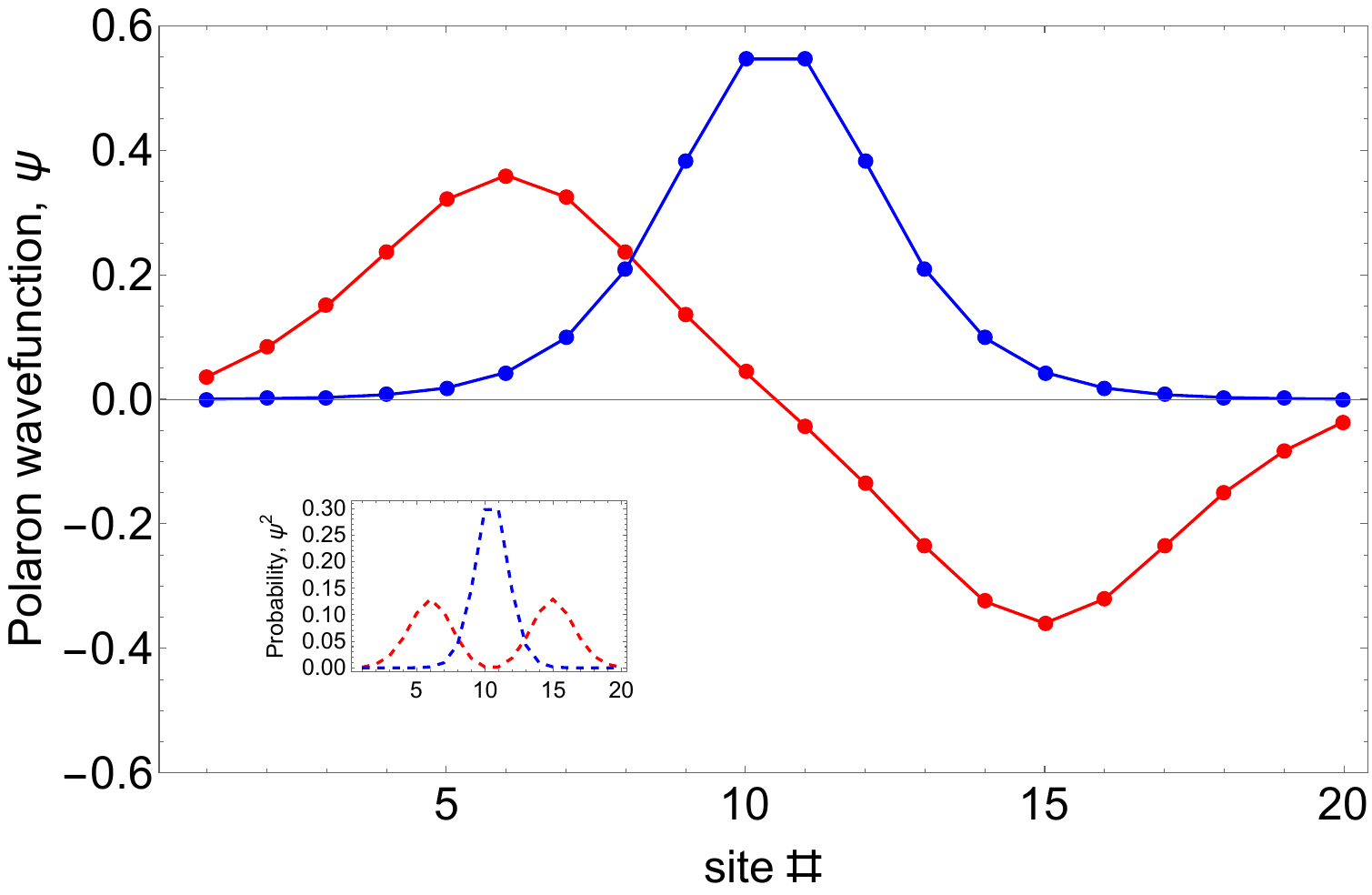}
    \caption{Ground state (blue) and excited state (red) polaron wavefunctions $\Psi$ on straight P3HT chain consisting of 20 monomers. 
    The inset shows the probability distribution $\Psi^2$ 
    the same polaron wavefunction.}
    \label{fig:ExcitedPolaron_20}
\end{figure}

Using the tight-binding polaron model, 
we compute the ground- and excited-state 
polaron wavefunctions for P3HT chains of varying lengths. 
As shown in Fig.\ \ref{fig:ExcitedPolaron_10} 
\& \ref{fig:ExcitedPolaron_20}, 
the ground state is symmetric and localized, 
while the excited state is antisymmetric.
The corresponding ground- and excited-state energies 
decrease with increasing chain length 
and approach saturation values.

\begin{figure}[h!]
    \centering
    \includegraphics[width=0.75\linewidth]{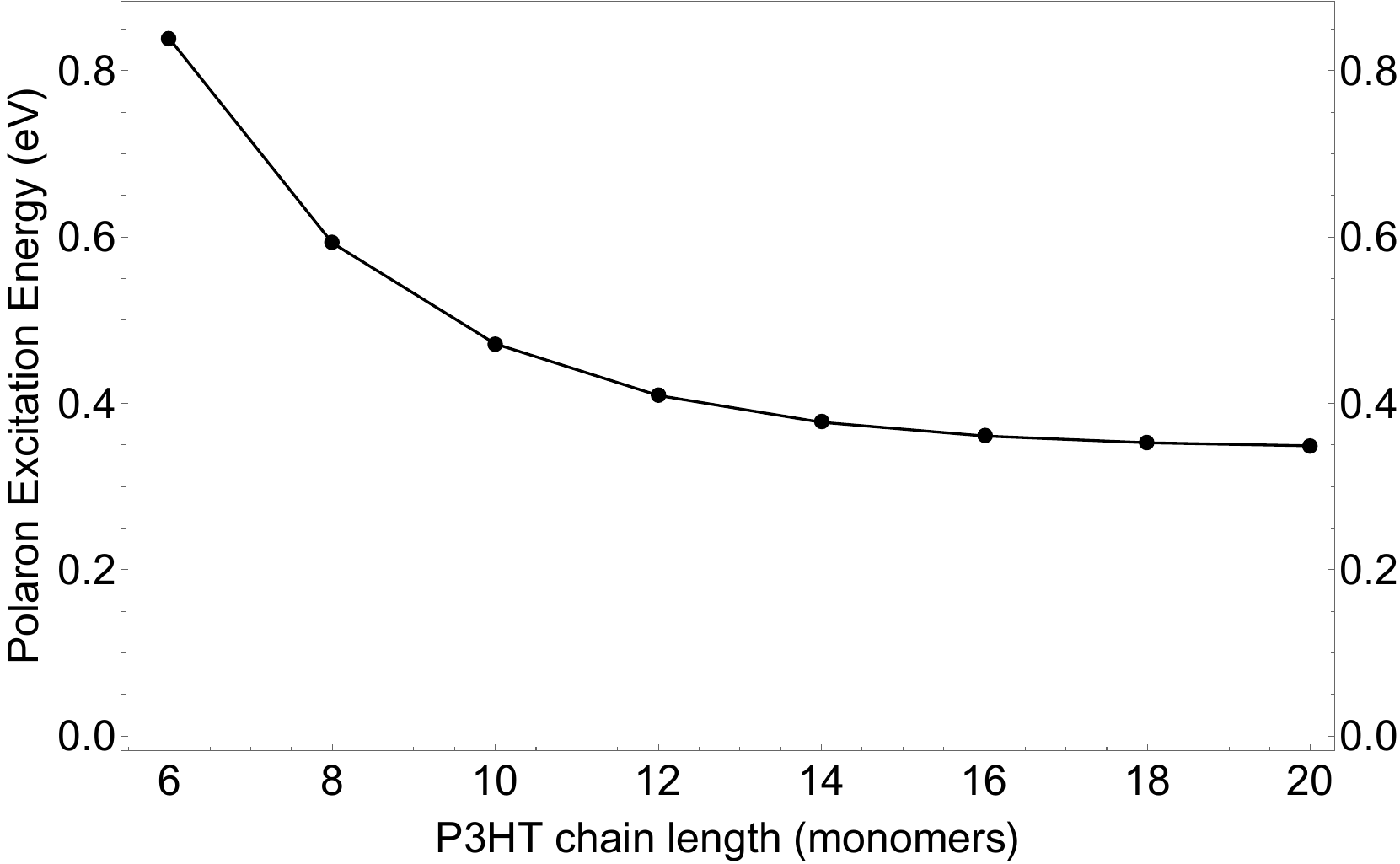}
    \caption{Polaron excitation energy as a function of oligothiophene (P3HT chain segment) length 
    in terms of monomers, for $\epsilon_r = 2.5$.}
    \label{fig:PolExEvsLen}
\end{figure}

\begin{figure}[h!]
    \centering
    \includegraphics[width=0.75\linewidth]{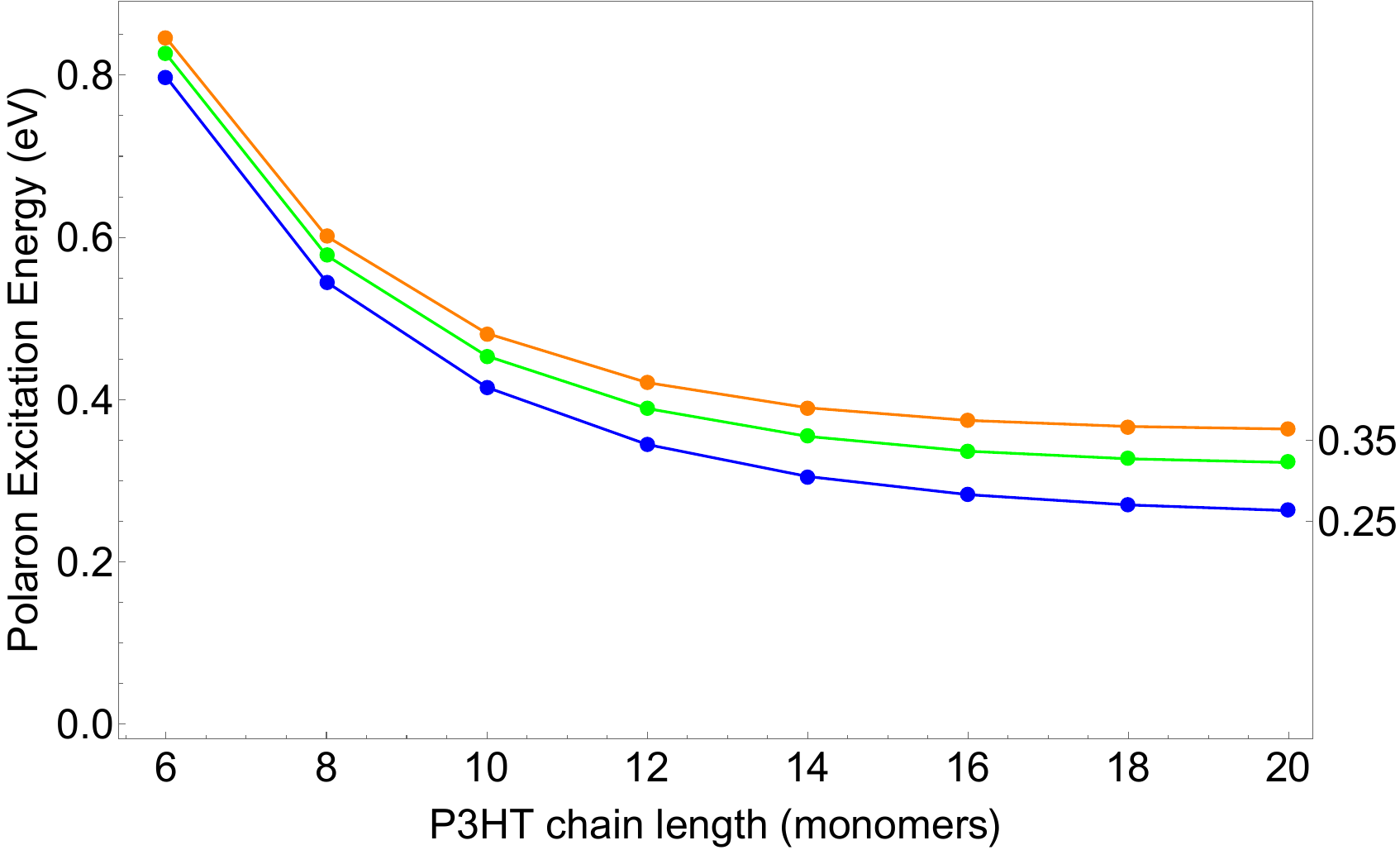}
    \caption{The polaron excitation energies as a function of P3HT chain length
    for (a) $\epsilon_r=2.0$ (blue), (b) $\epsilon_r=2.5$ (green) 
    and (c) $\epsilon_r=3.0$ (orange). }
    \label{fig:PolEE_sens}
\end{figure}

Fig.~\ref{fig:PolExEvsLen} shows the polaron excitation energies
as a function of P3HT chain length. 
The chain length reflects the number of monomers 
available for a polaron to delocalize freely.
For an $\epsilon_r = 2.5$, the polaron excitation energy decreases 
from $\sim 0.6$ eV for short chains containing about 8 monomers, 
to $\sim 0.3$ eV for longer chains containing 20 or more monomers.
Fig.\ \ref{fig:PolEE_sens} shows the sensitivity of the polaron excitation energy
with respect to the dielectric constant $\epsilon_r$ for P3HT.

Regioregular (RR) P3HT consists of long, planar chains 
that permit extended polaron delocalization, 
whereas regiorandom (RRa) P3HT contains torsional disorder 
and kinks that limit the conjugation length. 
The calculated excitation energies account for the shift
in the intrachain polaron absorption peak (peak B) 
from approximately $0.25-0.35$ eV for RR-P3HT 
to $0.5-0.6$eV for RRa-P3HT,
consistent with experimental observations shown in Fig.~\ref{fig:ExpPolAbsSpec}.

The density of polaron states in amorphous P3HT exhibits a standard deviation of about 0.2 eV, 
which explains the pronounced broadening of the polaron absorption 
spectra for regiorandom P3HT 
(blue Peak B in Fig.\ \ref{fig:ExpPolAbsSpec}).
In contrast, the greater molecular order and crystallinity in regioregular P3HT
lead to narrower distribution of polaron energies, 
and a smaller broadening around peak B for RR-P3HT.

\section{Alternative mechanism: ring distortion stabilization}

While the dielectric-stabilized tight-binding model successfully reproduces the experimentally observed polaron excitation energies across varying chain lengths, it is important to consider whether other physical mechanisms might also contribute to polaron stabilization. In particular, local geometric distortions such as ring deformations within the thiophene backbone have been proposed as a possible source of binding energy in conjugated polymers.

These distortions arise from the coupling between the charge carrier and vibrational modes of the polymer chain, and are often modeled using Holstein-like frameworks. Although such mechanisms are inherently local and short-ranged, they may still influence the shape and energy of polaronic states, especially in disordered or short-chain environments.

\begin{figure}[htbp]
\begin{center}
\includegraphics[width=0.5 \textwidth]{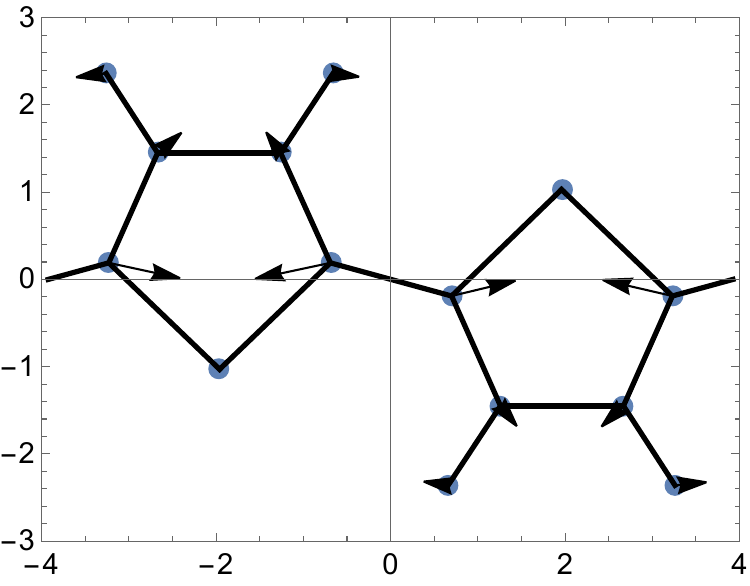}
\caption{Distortion of all-trans polythiophene dimer,
bonded to itself across the periodic boundary condition,
in response to a cation.}
\label{fig:distort}
\end{center}
\end{figure}

To assess the viability of this alternative, 
we now construct a simplified model of ring distortion stabilization.
When a cation sits on a polythiophene chain,
its rings distort slightly in response.
Fig.\ \ref{fig:distort} shows the atomic displacements
(multiplied by a factor of 15) 
for an all-trans polythiophene dimer,
bonded to itself through the periodic boundary condition,
calculated using the quantum chemistry package CP2K (see below for details).
For the periodic dimer, 
the cation sits with equal probability on both rings,
and the ring distortion is identical on both rings.

The ring distortion of Fig.\ \ref{fig:distort}
does not correspond precisely to a single vibrational mode,
but is instead the weighted sum of contributions from several modes.
Here, we simplify the multimode problem
by representing many ring modes as a single effective mode,
with an eigenvector defined by the displacement induced by a cation.
The effective mode has a frequency $\omega$,
with which the molecule would oscillate 
if it were constrained only to distort along the given eigenvector.

We then write the quantum Hamiltonian in terms 
of creation and annihilation operators,
with a linear coupling term corresponding 
to a constant displacement force:
\begin{equation}
H = \hbar \omega \left( a^\dag a + 1/2 \right) 
- \lambda f \left( a + a^\dag \right)
\label{eq:ham}
\end{equation}
Written in terms of oscillator raising 
and lowering operators $a^\dag$ and $a$,
the position operator $q$ is proportional to $a + a^\dag$,
so we write the coupling term in this form.
Physically, the displacement force $f$ arises
from the presence of a carrier (cation or anion) on the ring.
We choose to represent the force 
by the dimensionless probability $f$ that a carrier is present,
so that $\lambda$ has units of energy. 

The coupling term displaces the oscillator from its ground state
by some amount $\Delta x$ proportional to $f$.
We determine the coupling constant $\lambda$
by finding the displaced ground state
that minimizes the expectation value of the Hamiltonian Eqn.\ \ref{eq:ham},
and comparing its properties to DFT calculations.


In terms of the momentum and position operators $p$ and $q$,
the displacement operator $T(\Delta x)$ is
\begin{equation}
T(\Delta x) = e^{-i p \Delta x/\hbar}
\end{equation}

The operator $T(\Delta x)$ applied to any position eigenstate $|x \rangle$
has position eigenvalue $x + \Delta x$.

The raising and lowering operators $a^\dag$ and $a$
are related to $p$ and $q$ by 
\begin{eqnarray}
q &=& \sqrt{\frac{\hbar}{2 m \omega}} \left( a + a^\dag \right) \nonumber \\
p &=& i \sqrt{\frac{\hbar m \omega}{2}} \left( a^\dag - a \right) 
\label{eq:pqa}
\end{eqnarray}

From these relations it can be verified that $a$ and $a^\dag$
have the canonical commutation relation
$ \lbrack a, a^\dag \rbrack = 1 $
and that the oscillator Hamiltonian can be written as
\begin{equation}
H = \frac{p^2}{2m} + \frac{kq^2}{2} = \hbar \omega \left( a^\dag a + 1/2 \right)
\end{equation}
with $\omega^2 = k/m$.


Given Eqn. \ref{eq:pqa} for $p$ in terms of $a$ and $a^\dag$,
we can write the displacement operator $T(\Delta x)$ as 
\begin{equation}
T(\alpha) = e^{\alpha a^\dag - \alpha^* a}
\label{eq:coh_disp}
\end{equation}
with $\alpha$ defined as
\begin{equation}
\alpha = \Delta x \sqrt{\frac{m \omega}{2 \hbar}} 
\label{eq:alpha_dx}
\end{equation}

An operator $T(\alpha)$ of the form Eqn.\ \ref{eq:coh_disp} 
applied to the ground state $|0 \rangle$
generates a ``coherent state'' $| \alpha \rangle$:
\begin{equation}
T(\alpha) |0 \rangle = | \alpha \rangle
\end{equation}
which is an eigenstate of the lowering operator $a$:
\begin{equation}
a | \alpha \rangle = \alpha | \alpha \rangle
\end{equation}
Coherent states evolve with the same equations of motion
as the corresponding classical oscillator states.
$T(\alpha)$ with $\alpha$ purely real (as in Eqn.\ \ref{eq:alpha_dx}) 
generates a ``displaced state''.

Now we evaluate the expectation value of the energy
in the state $\alpha$:
\begin{equation}
\langle \alpha | H | \alpha \rangle = \hbar \omega (\alpha^2 + 1/2) - 2 \lambda f \alpha
\end{equation}
Minimizing with respect to $\alpha$, we find
\begin{equation}
\alpha = \frac{\lambda f}{\hbar \omega}
\label{eq:alpha}
\end{equation}
The corresponding minimum energy of the displaced oscillator is
\begin{equation}
\langle \alpha | H | \alpha \rangle = E_0 - \frac{(\lambda f)^2}{\hbar \omega}
\end{equation}
Aside from the zero-point contribution $E_0 = \hbar \omega/2$,
the energy is negative,
equal to half the stabilization energy $-2\lambda f \alpha$.

\textbf{Determining parameters}

We want to fit the oscillator frequency $\omega$,
coupling constant $\lambda$, and value of $\alpha$
for a distorted ring with a single charge.
From DFT calculations, we have 
the energy of the undistorted ground state $G_0$,
the energy of the optimized hole $H_H$,
the energy of the ground state in the hole configuration $G_H$,
and the energy of the hole in the ground state configurations $H_0$.
We can interpret the following quantities:
\begin{itemize}
\item $H_0 - G_0$ is the hole self-energy;
\item $G_H - G_0$ equals $\hbar \omega \alpha^2$,
the ring elastic distortion energy; and
\item $H_H - H_0$ equals $\hbar \omega \alpha^2 - 2 \alpha \lambda f$,
the ring distortion plus stabilizing energies.
\end{itemize}

Making use of Eqn.\ \ref{eq:alpha}, we have 
\begin{eqnarray}
G_H - G_0 &=& \frac{(\lambda f)^2}{\hbar \omega} \nonumber \\
H_H - H_0 &=& - \frac{(\lambda f)^2}{\hbar \omega} 
\label{eq:lambda}
\end{eqnarray}
(Below, we verify that these two computed values are indeed equal and opposite.)
If we knew $\omega$, we could determine $\lambda f$ and hence $\alpha$.

To determine the effective frequency,
we make use of the fact that we know 
the displacements $\{r_i\}$ and masses $\{m_i\}$ 
of all the atoms in the ring
in response to the presence of a hole,
with corresponding energy $G_H - G_0$.
If the effective mode oscillates in time,
the atoms would move according to
$r_i(t) = r_i \cos (\omega t)$.
The average potential energy and average kinetic energy
for a harmonic oscillator are equal,
and hence both are equal to half the total energy, $(G_H - G_0)/2$.

We can write the average kinetic energy explicitly as
\begin{equation}
K = \langle \sum_i (1/2) m_i r_i^2 \omega^2 \cos^2 (\omega t) \rangle
\end{equation}
Performing the time average and rearranging, we have
\begin{equation}
\omega^2 = \frac{G_H - G_0}{\sum_i (1/2) m_i r_i^2}
\label{eq:freq}
\end{equation}

We get the displacements $\{ r_i \}$ 
from the CP2K {\tt .xyz} output files
for the optimized ground and excited states.
Because these calculations were done 
for periodically self-bonded thiophene dimers, 
we have $f=1/2$.

The four energies $G_0$, $H_0$, $G_H$, and $H_H$ were computed using 
CP2K with a 400 Ry cutoff and the Pade exchange-correlation functional. 
TZV2P-GTH basis sets and GTH-PADE pseudopotentials were used for 
H, C, and S atoms to model their electronic structure accurately.
The results are given in Table 1,
where the values reported are the energies per ring 
(i.e., we have divided the total dimer energy by two;
likewise, because the hole sits equally on both rings,
in calculations that follow we take $f=1/2$.)
\begin{table}[htbp]
\begin{center}
\begin{tabular}{c|c}
State & Energy (H) \\
\hline
$G_0$ & -34.1606 \\
$H_0$ & -34.1064 \\
$G_H$ & -34.1556 \\
$H_H$ & -34.1113 
\end{tabular}
\end{center}
\label{tab:4energies}
\caption{Four fundamental energies for periodic thiophene dimer.}
\end{table}%

We observe that $G_H - G_0 = 0.005$H
is nearly equal and opposite to $H_H - H_0 = -0.0049$H,
as expected from our stabilization analysis. 

The moment of inertia $I = \sum_i (1/2) m_i r_i^2$,
computed from the atomic displacements 
from optimized ground state to optimized cation state,
equals 0.0392 g \AA$^2$/mol.

The effective frequency computed from Eqn.\ \ref{eq:freq}
equals $1.830 \times 10^{14}$ Hz,
which corresponds to an energy of 0.120 eV.
With $f=1/2$ for the periodic dimer cation,
from Eqn.\ \ref{eq:lambda} we find $\lambda = 0.256$ eV.
From Eqn.\ \ref{eq:alpha}, we then find $\alpha=1.063$.

\textbf{Polaron excited states}

We seek to compare the strength of polaron stabilization 
from ring distortion to that from dielectric polarization.
We can do this by computing a polaron stabilized by ring distortion alone,
and comparing the resulting stabilization energy 
and wavefunctions to results for the dielectric polaron.

The tight-binding Hamiltonian to describe 
a ring-stabilized polaron on a single straight chain 
takes the form
\begin{equation}
H = -2t \sum_k \psi_k \psi_{k+1} - c \sum_k |\psi_k|^4
\end{equation}

The first term is the one-body kinetic energy of the cation,
with hopping matrix element $t$ between adjacent sites
(the zero of kinetic energy here is a cation localized on one site).
The second term is the ring-mode stabilization,
with $|\psi_k|^2 = f_k$ the probability 
for the cation to sit on site $k$,
and coupling constant $c = \lambda^2/(\hbar \omega)$.
With values from the previous section, $c = 0.544$ eV.

We compute polaron properties by numerically minimizing 
the expectation value $\langle \psi | H | \psi \rangle$
with the constraint that $\psi$ is normalized, 
$\langle \psi | \psi \rangle = 1$.
Our initial guess for $\psi$ is a Gaussian
centered at the middle of a 29-site array,
with a width $\sigma$ such that $1/(2 \sigma^2) = 0.2$,
i.e., $\sigma = \sqrt{5/2} = 1.58$.

\begin{figure}[htbp]
\begin{center}
\includegraphics[width=0.7 \textwidth]{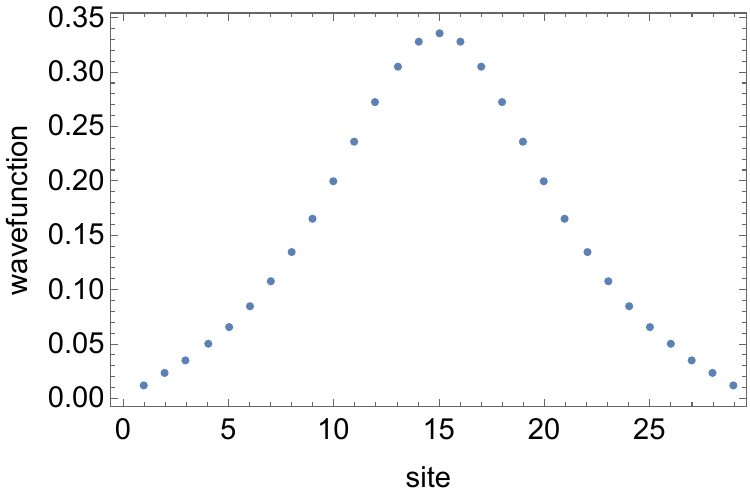}
\caption{Polaron wavefunction for cation on all-trans polythiophene.}
\label{fig:ground}
\end{center}
\end{figure}

Assuming a $t$ value of 1.3 eV,
the resulting polaron has a mean-square width
of $\sigma = 3.8$ sites,
and an energy of -2.618 eV.
Fig.\ \ref{fig:ground} displays 
the optimized ground-state wavefunction.

Recall that this energy is with respect to the fully localized state;
a fully extended state for the cation would have energy $-2t = -2.6$ eV,
so the polaron binding energy with respect to the fully extended state
is only 0.018 eV, less than the thermal energy kT $=0.025$ eV.
This ring-stabilized polaron is extremely weakly bound
compared to the dielectric polaron computed previously.

The unimodal ground-state wavefunction 
is not the only possible polaronic state.
We can compute a succession of excited states
by enforcing orthogonality with respect 
to all previously computed wavefunctions.
This is straightforward for the first excited state;
because the ground state is even 
under reflection about its center,
the first excited state must be odd.

\begin{figure}[htbp]
\begin{center}
\includegraphics[width=0.7 \textwidth]{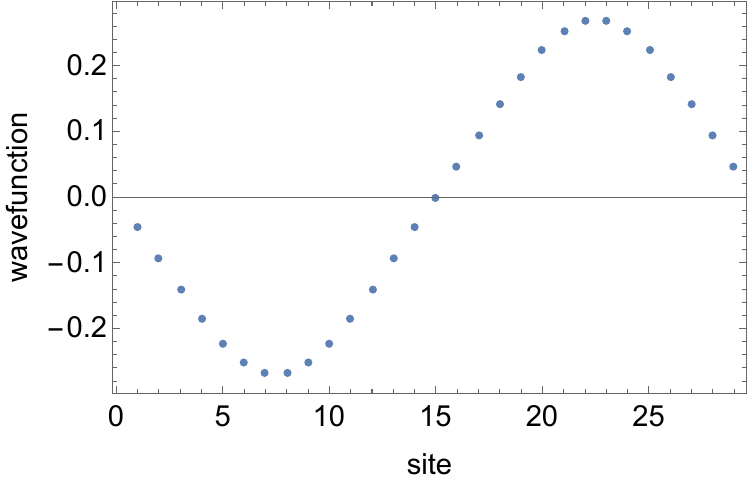}
\caption{First excited state ring-stabilized polaron wavefunction on polythiophene.}
\label{fig:excited}
\end{center}
\end{figure}

We can find the first excited state
by minimizing the expected energy
with respect to a wavefunction
that is enforced to be odd by construction,
in which only values for the left half 
of the wavefunction are minimized over.
Our initial guess is a discretized version
of the function $x e^{-x^2/(2 \sigma^2)}$,
centered on the middle site as before,
with a larger width $\sigma=\sqrt{10} = 3.16$.
Fig.\ \ref{fig:excited} shows the minimized wavefunction.

The difference in energy between the excited and 
ground state polaron is extremely small, only 0.047 eV,
which corresponds to an infrared photon of 380 cm$^{-1}$.
This value is much smaller than the range of energies
identified by spectroscopists as corresponding to polaron excitations.
This discrepancy is strong evidence that ring distortions
cannot be the main stabilization mechanism for polarons on P3HT.
This mechanism yields much weaker stabilization and significantly 
lower excitation energies which are inconsistent with experimental spectra.

\section{Discussion}

Polaron excited states in conjugated polymers are 
directly accessible via absorption spectroscopy, 
and these spectral features provide a sensitive test 
for polaron stabilization mechanisms. 
By comparing the predicted excitation energies 
with experimental absorption spectra, 
one can assess whether dielectric polarization 
or local structural distortions dominate polaron formation.

We applied a dielectric-stabilized tight-binding model 
to compute polaron ground and excited states in P3HT.
The model quantitatively reproduces the experimentally observed 
polaron excitation energies as a function of chain length, 
with excitation energies saturating near 0.3~eV for long chain segments 
and increasing to $\sim$0.5-0.6 eV for shorter conjugated segments. 
This chain-length dependence directly explains the observed shift 
in intrachain polaron absorption peak B between regioregular and regiorandom P3HT films. 
The agreement between theoretical predictions 
and experimental data validates dielectric stabilization 
as the dominant mechanism governing polaron formation in P3HT.

In contrast, we examined an alternative stabilization mechanism 
based on local ring distortions within the thiophene backbone. 
Using quantum-chemical calculations to parameterize a Holstein-like model, 
we find that ring distortions contribute only weak stabilization, 
yielding binding energies below thermal energy ($k_B T$) and 
polaron excitation energies that are much smaller than those observed experimentally. 
This demonstrates that local structural distortions have only 
a weak contribution to polaron stabilization and 
are insufficient to reproduce the spectroscopic trends observed in P3HT.


\bibliographystyle{unsrt}
\bibliography{references.bib}

\end{document}